\documentclass[final,3p,times,twocolumn,authoryear]{elsarticle}

\usepackage{amssymb}
\usepackage{lipsum}

\usepackage{float}
\usepackage{enumitem}
\usepackage{textcomp}
\usepackage{mathtools}
\usepackage{commath}
\usepackage{float}
\usepackage{graphicx}
\usepackage{epstopdf}
\usepackage{caption}
\usepackage{hyperref} 
\usepackage{tocloft} 
\usepackage{fancyhdr} 
\usepackage{lastpage}
\usepackage{sidecap}
\usepackage[table,xcdraw]{xcolor}
\usepackage{subcaption}
\usepackage{indentfirst} 
\usepackage{tabularray}
\usepackage{changes}
\usepackage{gensymb}
\definechangesauthor[name=diego, color=orange]{D}

\begin{document}

\begin{frontmatter}



\title{Snow cover over the Iberian mountains in km-scale global climate simulations: evaluation and projected changes.}

\author[UCM]{Diego García-Maroto}
\author[UCM]{Elsa Mohino}
\author[UCM]{Luis Durán}
\author[UCM]{Álvaro González-Cervera}
\author[ECMWF]{Xabier Pedruzo-Bagazgoitia}

\affiliation[UCM]{organization={Departamento de Física de la Tierra y Astrofísica, Universidad Complutense de Madrid},
            city={Madrid},
            postcode={28040} ,
            country={España}}

\affiliation[ECMWF]{organization={European Centre for Medium-Range Weather Forecasts (ECMWF)},
            city={Bonn},
            country={Germany}}

\begin{abstract}

Mountains represent a crucial natural resource for many regions worldwide, particularly in Mediterranean areas such as the Iberian Peninsula, where climate change can have profound consequences. However, due to their coarse resolution, traditional climate models cannot properly represent the complexity of orographic processes. Storm-resolving models, with horizontal resolutions finer than 10 km, offer a path forward to provide high-quality information for local adaptation even in complex terrain. In this work, we assess the performance of one such model, the IFS-FESOM developed within the EU-NextGEMS project, in simulating the historical seasonal snow cover in the main Iberian mountain ranges and projecting climate change effects. It is evaluated against four high-resolution reanalysis-based products, satellite, and in situ observations. Despite a positive bias in the length of the snow cover season and the number of snowfall days, the model can represent features of the snow climatology, such as its elevation dependency, seasonal cycle, and mean covered area, successfully within the range of uncertainty of the reference data. In addition, the projection under the SSP3-7.0 scenario shows a marked reduction of the snow season across almost all of Iberia, with low- to mid-elevations and the southernmost ranges showing the greatest reductions. These changes are linked to reduced snowfall days, caused primarily by rising temperatures but also by decreasing precipitation, particularly in southern and Mediterranean sectors. Our results suggest that storm-resolving simulations can provide local climate projections of snow information in complex terrain without regional downscaling.

\end{abstract}




\end{frontmatter}



\section{Introduction} 
Mountain regions play a fundamental role in the hydrological cycle of vast regions of the world \citep{water-tower1,water_tower2}, largely due to local processes such as orographic precipitation and the presence of seasonal or permanent snow cover. In the context of climate change, it is anticipated that some of these processes will be subject to disruption, resulting in substantial impacts on local ecosystems and nearby populations \citep{med_water_2011,pepin_elevation-dependent_2015}. This is particularly relevant for regions such as the Iberian Peninsula, where the development of a seasonal winter-spring snowpack confined to various medium-sized mountain ranges is key to offsetting water deficits during the dry summer season that characterizes its Mediterranean climate \citep{med_pyrenees_2008,med_mountains_hidro_2009,med_mountains_europe_2018,alonso-gonzalez_snow_2020}. It is therefore vital to ascertain the future climate of these mountains, and more specifically, to project possible changes to the seasonal snowpack. This is of particular importance for the management of water resources, and also for a wide range of economic interests, including tourism, logistics, infrastructure management, and risk mitigation \citep{med_mountains_cc_2008, med_water_2011}.

The Intergovernmental Panel on Climate Change (IPCC) Special Report on the Ocean and Cryosphere in a Changing Climate \citep{ipcc-special-mounts} reports a worldwide decline in snow cover, especially pronounced at low elevations. These changes have been identified to affect the seasonality of runoff and have therefore been found to impact water availability and agriculture. Reduced seasonal snow cover has also been reported to be responsible for habitat changes and species redistribution in high mountain areas \citep{species_vegetation_cs, ranas-luis,species_trees2_2008,species_forest_forest_2012,species_trees_med_mountain, especies, habitat_alps, viescopausa,species_fire_2023}, as well as having negative impacts on the reproductive cycle of some snow-dependent species \citep{renos, especies}. The IPCC Special Report also cites the reduction of snow cover as a challenge for winter tourism, especially for the viability of operating low-elevation ski resorts. In almost all regions, snow cover is projected to continue its decline throughout the 21st century. In this context, efforts are necessary in order to anticipate the future challenges at the regional to local levels so that adaptation action can be carried with sufficient information. 

Although standard state-of-the-art global climate models, as the ones participating in the coupled model intercomparison project phase 6 \citep[CMIP6,][]{eyring2016overview}, can be successfully used to simulate recent large-scale trends in snow cover in the Northern Hemisphere \citep{ipcc-special-mounts}, their coarse resolutions, typically of $\sim$100 km, are insufficient in providing regional data, especially at medium-sized mountain ranges such as those within the Iberian Peninsula. For this reason, to achieve a useful representation of orographic processes and to gather richer information for regional research and adaptation, downscaling efforts are usually necessary. Statistical downscaling approaches have been widely applied to infer local-scale mountain climate from large-scale predictors \citep{RASCAL,HICAR-COSD,downscaling_AI_2024}. However, these approaches are empirical, thus often dependent on the existence of accurate observations, which are scarce in mountain areas. Furthermore, snow accumulation and persistence are controlled by highly nonlinear, small-scale mechanisms such as orographic precipitation, wind redistribution, and shading \citep{snow_distrib_heter_1991,snow_distrib_heter_1992,snow_distrib_het_2011,snow_distrib_het_review_2011,HICARsnow}. Consequently, statistical methods often fail to reproduce the spatial heterogeneity and elevation gradients that dominate mountain snowpack \citep{HICAR-COSD}. For this reason, physically based dynamical downscaling remains traditionally the most suitable approach to represent snow processes at regional scales. These efforts rely either on the development of specific modeling targeted at the area and problem in mind \citep{downscaling_2013,duran2018multi, ICAR, downscaling_2021, HICAR, HICARsnow} or on international efforts  such as the CORDEX initiative \citep{downscaling_CORDEX_2024}. This generates a significant time gap between the release of new global climate simulations and their use in the assessment of complex variables such as snow cover presence on local to regional scales. Moreover, dynamical downscaling approaches can present large-scale inconsistencies with respect to the global model from which boundary conditions were taken, which is unlikely to be a result of upscaled added value \citep{taranu2023mechanisms}.

To overcome some of these problems and achieve other long-awaited improvements in recent years, the global modeling community has been steadily moving into the idea of global km-scale modeling \citep{nextgems}. These models have advantages such as explicitly resolving oceanic eddies or even convection when resolution is fine enough, so they are often regarded as storm-resolving or eddy-resolving models. One of their main advantages is that with global km-scale modeling, high-resolution simulations can be obtained with a much simpler and shorter chain, as just one coupled model is enough. This also equips all regions, as global simulations allow the study of regional features without requiring specific resources for downscaling a particular area. Another advantage is that without a chain of models, interaction across different scales and regions is possible, so regional climate impacts on global large-scale circulation can be reproduced. 

\begin{figure*}[h!]
	\centering
	\includegraphics[width=1\textwidth]{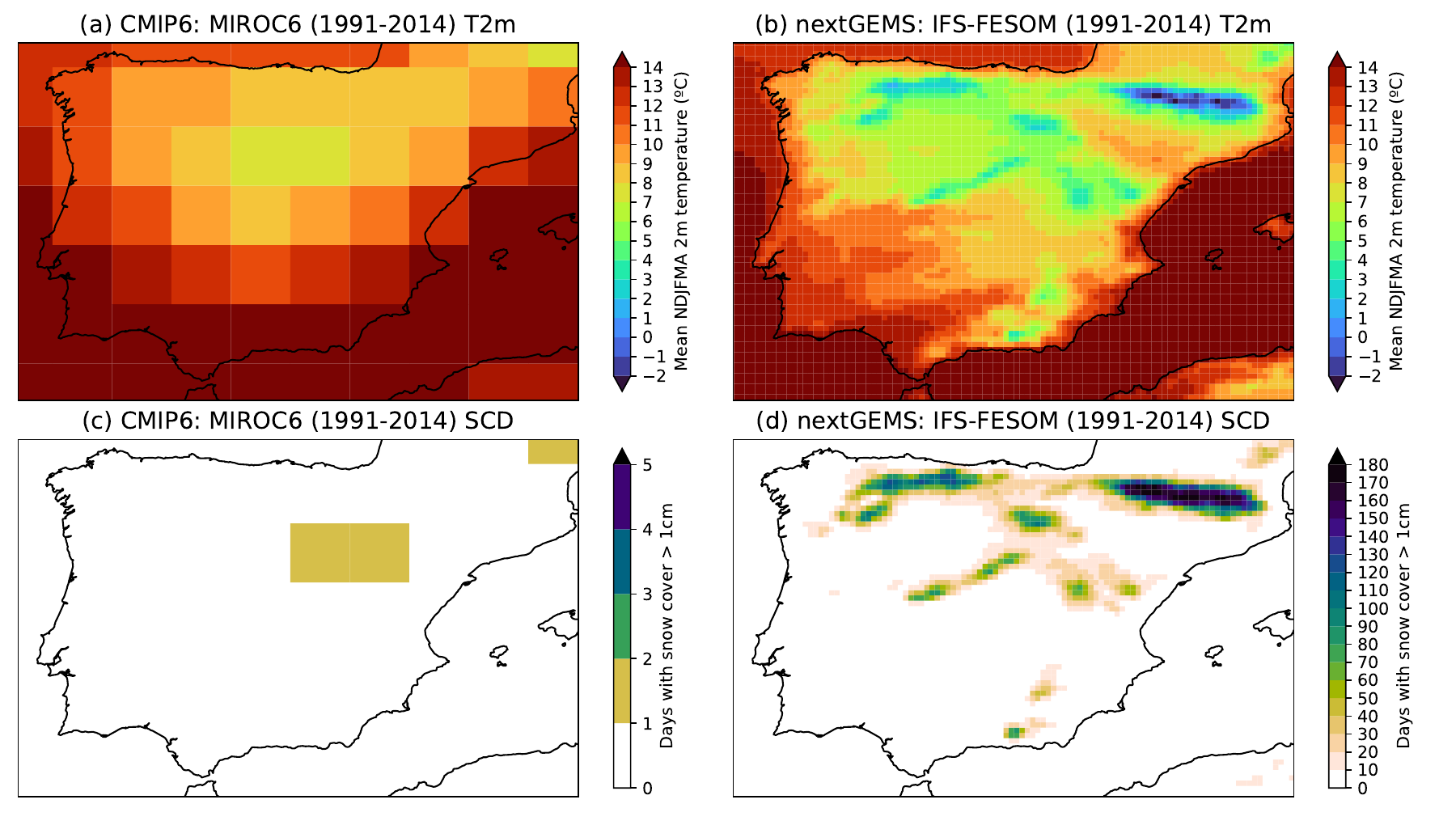}
	\caption{Comparison between historical simulations (1991-2014) of CMIP6 model MIROC6 (a,c) and nextGEMS IFS-FESOM model (b,d). Mean extended winter (NDJFMA) 2-meter temperature (a,b) and number of days with a snow cover over 1 cm (c,d). Note that c and d feature different color-bar scales.}
	\label{fig: intro_nextGEMS_vs_CMIP6}
\end{figure*}

One of the first projects aiming to develop such km-scale models and the first multidecadal km-scale simulations is H2020 Next Generation of Earth Modeling Systems (nextGEMS) \citep{nextgems}. Within the project, 30-year-long simulations at 9km resolutions were produced with two models: ICON and IFS-FESOM. In particular, the IFS-FESOM model was used to produce a 1990 to 2019 historical simulation and a 2020 to 2049 SSP3-7.0 projection, allowing for both verification and climate change assessment. As noted, these simulations allow the direct study of complex regional climate features, such as seasonal snow cover in medium-sized mountains like those of the Iberian Peninsula, which is not possible with coarser CMIP6 models unless subjected to regionalization. Figure \ref{fig: intro_nextGEMS_vs_CMIP6} presents an example of one of such CMIP6 models, the MIROC6  \citep{tatebe2019description}, where horizontal resolution is too coarse to resolve any zones with mountain climate characteristics in Iberia. Conversely, the km-scale resolution of the nextGEMS IFS-FESOM model provides a much more accurate representation of winter temperature and its dependency on the orography, together with a more realistic representation of snow cover duration (Figure \ref{fig: intro_nextGEMS_vs_CMIP6}). 

This work has two main objectives: Firstly, we aim to verify whether the km-scale global multidecadal simulations performed with the IFS-FESOM model can correctly represent the climatology of the seasonal snowpack present in the main Iberian mountain ranges, while also identifying possible limitations or biases. To do so in a region characterized by a scarcity of snow observations, we use manned station data, together with reanalysis, gridded, and satellite products. Our second aim is to assess the near-term projection of climate change in these vulnerable areas, and determine the key drivers of such changes at the regional to local level.

The article is structured as follows. Section 2 describes the IFS simulations as well as each of the reference reanalysis-based datasets, manned and satellite observations. Section 3 defines the studied areas and describes the pre-procesing, index calculations and statistical methods employed. Section 4.1 focuses on the first objective of evaluating the performance of the model at representing different features of the snow cover season such as its length, seasonal cycle and interannual variability. Biases are also identified and discussed. Section 4.2 focuses on the second objective by computing changes between the historical and projection runs and discussing their possible drivers. Finally, the main results are summarized and conclusions are drawn in section 5.  

\section{Simulations and reference data}
\subsection{IFS-FESOM simulations}
Within the nextGEMS project, various multi-decadal km-scale global climate simulations were developed \citep{nextgems}. Two simulations from 2020 to 2049 following a SSP3-7.0 scenario \citep{o2016scenario} were generated, one using the ICON model and the other using IFS-FESOM (hereafter IFS) \citep{ifs-nextgems}. In addition, the IFS was also used to produce a historical simulation from 1990 to 2019. In this work, we only considered the simulations performed by the IFS model for two main reasons. Firstly, the historical IFS simulation allows for a validation against observations. Secondly, the snow depth water equivalent and snow density variables necessary for our analysis were only available for the IFS runs. 

The Integrated Forecasting System (IFS) is developed and maintained by the European Centre for Medium-Range Weather Forecasts (ECMWF) together with its member states. It approaches these simulations from the philosophy of an operational numerical weather prediction model; thus, it is equipped with elaborated parameterizations (including convection) and tuning options \citep{nextgems}. The simulations analyzed here use a weakly active convection scheme \citep{ifs-nextgems}. The atmosphere and land modules use the octahedral-reduced Gaussian (TCo) grid and run at a resolution of about 9 km, while the ocean uses the FESOM2.5 model and an eddy-resolving grid at up to 5 km resolution (NG5) \citep{ifs-nextgems}. The land surface uses the ECLand model \citep{ecland}, which has been equipped with a multi-layer snow scheme \citep{snow_ifs} that provides snow characteristics such as depth, density, liquid water content, and albedo as prognostic variables \citep{ifs-nextgems}. Both simulations were performed in the native grid combination TCo1279-NG5 and then transformed and saved into a HEALPix (Hierarchical Equal Area isoLatitude Pixelation) grid for homogenization with the ICON model. HEALPix grids allow for different zoom levels in order to easily change from one resolution to another, but the maximum zoom level at which the IFS simulations were saved was level 9 which corresponds to a resolution of about 12.5 km.

This work analyses the snowpack both from the historical and the SSP3-7.0 projection simulations. It also considers other, more general variables such as temperature, snowfall, or total precipitation to assess the possible origins of both biases and future changes detected in the snow depth variables. 

\subsection{Reference data}
For the main objectives of this work, the analysis focuses on the presence of ground snow layers exceeding a defined minimum thickness. Analyzing the presence of a snow cover is key for defining the cold season in mountain regions as it provides the start, end, and total number of days when the soil-atmosphere interaction is affected by snow, resulting in important interactions for the mountain ecology, such as changes in albedo and thermal insulation \citep{snow_heat_transfer, snow_albedo,snow_heat_transfer2}.

Despite the importance of understanding snow climatology, observations of snow depth and snow presence on the Iberian Peninsula remain extremely scarce. Some manual observations collected by staff from the Spanish State Meteorological Agency (AEMET) exist and provide valuable information for mountain studies, but they remain insufficient for comprehensive model validation. To mitigate this limitation, we use several publicly available reanalysis and gridded products that provide snow depth data for the region. Multiple products are employed instead of a single dataset to account for potential biases and inaccuracies in representing snow climatology, as identifying the best-performing product is also limited by the lack of ground observations. When available, satellite observations are additionally incorporated to complement these datasets and improve model verification, providing spatial regional context that point measurements alone cannot offer.

The reference data considered in this work for the validation of seasonal snow cover climatology in the IFS historical simulation is composed of four high-resolution gridded datasets of different characteristics, accompanied by satellite data and manual daily snow depth observations. For model validation, snow depth variables will be considered from all the different datasets. We describe each of the data sources in the following paragraphs. 
\subsubsection{ERA5-Land}
ERA5-Land \citep{era5land} is an operational product developed by the ECMWF within the Copernicus Climate Change Service (C3S) of the European Commission. It is a global dataset that enhances the land component of the fifth generation of European ReAnalysis (ERA5) covering the period from 1950 to the present at hourly frequency. Its main advantage over ERA5 is its enhanced horizontal resolution from the 31 km of ERA5 to 9 km. This is achieved by running the ECMWF land surface model (CHTESSEL) forced by downscaled near-surface atmospheric data from ERA5. The forcing is taken from the lowest ERA5 model level (10m above surface) and is interpolated from its original 31 km resolution to the ERA5-Land 9 km resolution via linear interpolation. It is important to note that ERA5-Land does not perform a direct data assimilation from observations, but these have previously been considered for the ERA5 atmospheric forcing that derived ERA5-Land. ERA5-Land has been found to improve the representation of multiple land surface features, such as soil moisture, lake description, and river discharge. Despite these, ERA5-Land snow variables have been found to have a mixed performance exhibiting different behaviors depending on location and altitude \citep{era5land,lei_snow_2023_era5land,kouki_evaluation_2023_era5land,varga_evaluation_2023_era5land,majidi_evaluation_2025_era5land}. It is also important to note that ERA5-Land employs a bulk-layer snow scheme, in contrast to the more advanced multi-layer snow scheme used by the IFS in nextGEMS \citep{snow_ifs}. 

\subsubsection{IPE-CSIC snow daily gridded dataset
}
\cite{ipe} presented and validated a daily gridded snow depth and snow water equivalent dataset for the Iberian Peninsula spanning the period 1980 to 2014 at a horizontal resolution of 10 km. The dataset was generated via a downscaling of the ERA-Interim reanalysis using the Weather Research and Forecasting (WRF) model to obtain the necessary high-resolution atmospheric forcing in order to drive a mass and energy balance snowpack model known as the Factorial Snow Model (FSM) \citep{fsm}. This model allows for multiple configurations on physics and parameterizations, thus permitting a high tuning for the simulations. The dataset has been validated and found to be fairly consistent with both satellite data and automatic snow depth observations \citep{ipe}.

\subsubsection{CERRA}
The Copernicus European Regional Reanalysis (CERRA) \citep{cerra} is a complete regional reanalysis of the entire Europe developed within the C3S spanning the years 1984 to 2021 at a horizontal resolution of 5.5 km. CERRA was produced using the HARMONIE (HIRLAM ALADIN Regional/Mesoscale Operational NWP In Europe) numerical weather prediction system. It provides both analysis data every 3 hours and short forecasts from 1 to 30 hours of lead time. The CERRA system uses ERA5 data as lateral boundary conditions and assimilates a large number of observations both in the atmosphere and on the surface via a 3D-Var scheme. CERRA assimilates both conventional observations (surface stations, buoys, radiosonde, aircraft measurements), satellite radiance data, and other non-conventional observations mainly from reprocessed datasets. CERRA has been validated and found to present an added value in most variables, especially in small areas with complex terrain such as the Alps \citep{cerra}. For our analysis, atmospheric fields from CERRA are also used to assess model biases in aspects other than snow presence.

\subsubsection{CERRA-Land}
In addition to CERRA, a specific land surface regional reanalysis was created within the C3S and named CERRA-Land \citep{cerra_land}. It shares spatial domain, time coverage, and horizontal resolution with CERRA but uses additional surface observations and more advanced parameterizations. It was produced by running the SURFEX (Surface Externalisée) V8.1 land surface model \citep{SURFEX} driven by atmospheric forcing from CERRA and an additional analysis of daily accumulated precipitation that combines information from CERRA and surface rain gauge observations. It uses an intermediate complexity multilayer snowpack scheme. 

\subsubsection{AEMET manual observations 
}
Although in situ observations of snow depth are scarce in the Iberian Peninsula and especially in its mountains, some monitoring sites do exist, and most are managed by the AEMET. The available data consists of daily manual observations recorded at certain fixed sites throughout the extended winter following the World Meteorological Organization (WMO) guidelines regarding snow depth measurements \citep{snow_measure_wmo}. 

The monitoring sites are spread throughout the Spanish territory, covering most of the main mountain ranges, but with important limitations. Many of the mountain sites are located in the Pyrenees, and fewer of them are on other important mountain ranges such as the Cantabrian Range. One exception is the Central System, with a robust and long record at Navacerrada station and Puerto de Cotos operated by  Sierra de Guadarrama National Park \citep{duran2017penalara}. Still, observations at elevations of more than 1500 m are mostly limited to the Pyrenees, with more low-elevation sites available for the other mountain ranges. Unfortunately, no data for the Sierra Nevada region was available within the series provided by AEMET. The time completion of the series is also to be taken into account: some are just a few years long or may barely overlap with the 1990-2019 historical period of interest for this work. 

Six representative sites were selected, considering at least two for each mountain range except for Sierra Nevada. The chosen sites were selected for the length of their complete time series within the 1990-2019 period, and to also sample sufficiently different altitudes and locations.

\subsubsection{Satellite data}

In addition to in situ measurements, satellite-derived snow products provide essential information for monitoring snow cover, especially in regions where ground observations are sparse, such as remote or high-altitude areas. In this study, we used the Normalized Difference Snow Index (NDSI) from the MODIS (Moderate Resolution Imaging Spectroradiometer) MYD10A1 (Aqua) dataset \citep{hall2016modis}, distributed by the National Snow and Ice Data Center. This product offers high temporal resolution, providing daily global snow cover maps at a native spatial resolution of approximately 500 m since 2003, and is therefore well suited for analyzing large-scale patterns in snow extent and seasonal variability \citep{hall2002modis, gascoin2015snow, saavedra2017snow}.

The NDSI data were processed to generate temporal series from 2003 to 2023 of snow presence at the grid points closest to the surface observation stations, as well as snow cover maps and snow-covered area (SCA) time series for each mountain range domain. The data were interpolated onto a regular latitude-longitude grid with a spatial resolution of $0.005^{^\circ}$. Each pixel was classified as snow covered when the NDSI exceeded the commonly used threshold of 0.4 \citep{dozier1989spectral}. The pixels classified as cloudy were removed and filled with snow or no-snow information using the method described by \citet{gafurov2009cloud}.

\section{Methods}
\subsection{Study areas}
\begin{figure}[h!]
	\centering
	\includegraphics[width=0.48\textwidth]{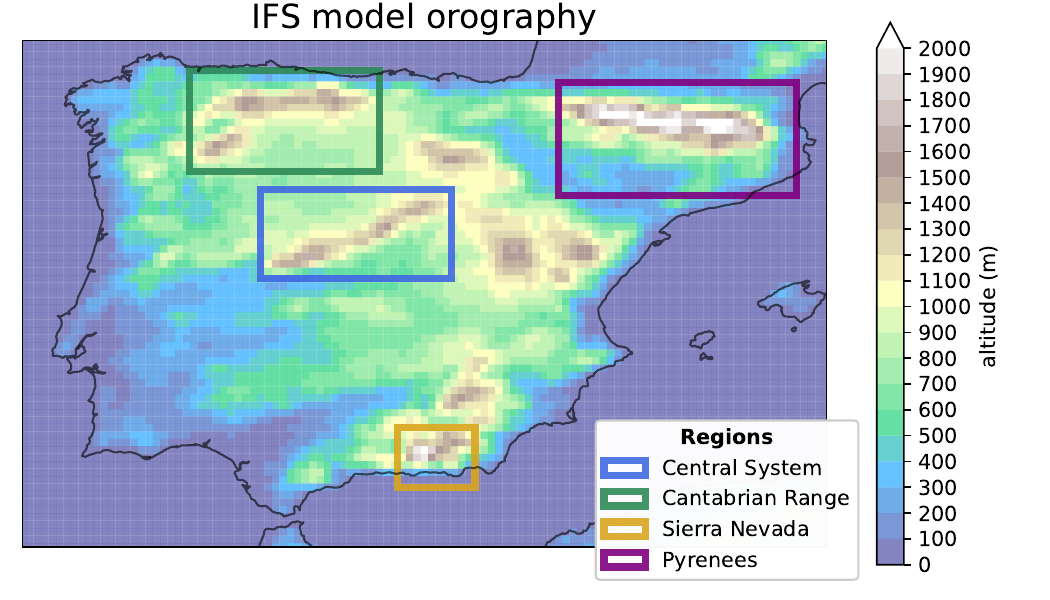}
	\caption{IFS-FESOM model orography interpolated to a regular 0.125º x 0.125º latitude-longitude grid. The four mountain regions that are studied in this work are highlighted, enclosed in rectangular areas that are later considered for analysis.}
	\label{fig: orog_ifs_zones}
\end{figure}

This work focuses on the main mountain areas of the Iberian Peninsula, which are the main places where a seasonal snow cover develops during the winter months in Iberia. Four mountain ranges have been selected for analysis based on their altitude, snow presence, and societal importance. These are: the Central System, the Cantabrian Range, Sierra Nevada, and the Pyrenees. Their location can be seen in Figure \ref{fig: orog_ifs_zones} along with the IFS model orography. Rectangular boxes have been considered around each of those mountain ranges for data selection and index calculations. 

\subsection{Data pre-processing}
To facilitate comparison between datasets, data has been pre-processed to homogenize the formats, units, and grids. 

The IFS output includes a variable labeled ``snow dept'', but this field actually represents snow water equivalent rather than the actual depth of the snowpack. Therefore, the true snowpack depth was computed as a quotient between snow water equivalent and snow density. The coarsest time resolution among the datasets is daily frequency (manual observations and the IPE-CSIC dataset). This is deemed sufficient for the calculation of different relevant snow indices, so all data stored at a higher frequency is aggregated into daily data via the median for snow depth, the sum for precipitation and snowfall, and the mean for other variables such as temperature.  

The datasets present different horizontal grids: the IFS simulations are in a zoom 9 HEALPix grid, ERA5-Land is in a regular 0.1º  latitude-longitude global grid, CERRA and CERRA-Land are in a Lambert conformal conical 5.5 km grid over Europe, and the IPE-CSIC dataset is in a rotated 10 km grid over the Iberian Peninsula. To facilitate inter-comparison, all datasets were initially linearly interpolated onto regular latitude–longitude grids with resolutions comparable to their respective native grids. Subsequently, when necessary, datasets with finer native resolutions were conservatively remapped to a common 0.125° × 0.125° latitude–longitude grid over the Iberian Peninsula. The resolution of this common grid is approximately equivalent to the 12.5 km resolution of the zoom 9 HEALPix, which is the coarsest of all the original grids. 

\subsection{Indices calculation}
For both model validation and projection analysis, several indices are computed. Except for the seasonal cycle, when monthly aggregation is used, the indices are calculated for the extended winter season within a water year (October to September), which is defined hereinafter as the months ranging from November to April, and noted as NDJFMA. This extended winter period covers almost all snow accumulation and snow ablation periods for the Iberian mountain ranges. We note that, as the simulations start and end in January instead of October, some months at the simulation edges are discarded, so that only complete extended winters are considered. The studied indices are detailed next:

\begin{itemize}
    \item Snow Cover Days (SCD): number of days in a given time period when snow depth was over 1 cm, calculated at each grid point.  
    \item Snow-Covered Area (SCA): geographical area where snow depth is over 1 cm. This is calculated by summing the number of covered grid-points multiplied by their latitude-weighted area and is defined for each day. Calculations are performed for each mountain range within its predefined bounding box (Figure \ref{fig: orog_ifs_zones}).
    \item Snowfall Days (SfD): number of days within a given period with at least 1 mm of accumulated solid precipitation, computed at each grid point.
    \item Frost Days (FD): number of days within a given period when minimum temperature was below 0 °C, computed at each grid point.
    \item Wet Day (WD): number of days within a given period when total precipitation exceeded 1 mm, computed at each grid point. 
\end{itemize}

\subsection{Multiple linear regression}
\label{sec: Methods_reg}
To identify possible drivers of the projected changes in snow cover duration and quantify their contributions at each grid point, we use multiple linear regression analysis \citep{libro_estadistica_Storch}: a response variable $Y$, such as the snow cover duration or other snow-related indices, could be explained by a linear combination of indices related to potential drivers. Denoting those drivers as $x_l$, one can express the multiple linear regression model without an intercept as: $Y_{ik} = \sum_{l=1}^{n} a_{lk} x_{lik} + \epsilon_{ik}$, where $a_{lk}$ are the fit coefficients, $i=1 ... T$ accounts for each extended winter, $k$ represents each simulated grid point and $\epsilon_{ik}$ are the residuals. As potential drivers, we test total precipitation and surface temperature. The model coefficients are estimated using the inter-annual variability provided within the 60 years of the historical and projection simulations, then the resulting fit is applied to assess the changes between the climatologies of the projection and historical simulations. Estimating the coefficients using only the historical simulation yields comparable results.

\subsection{Statistical significance}
In order to measure the statistical significance of the difference in mean climatology for both bias evaluation and assessment of future changes a two-sided t-test was performed. A Benjamini–Hochberg procedure was
then applied to the p-values to control the false discovery rate (FDR) ($\alpha_{FDR} = 0.05$) thus determining statistical significance while accounting for the multiple testing problem \citep{wilks_stippling_2016}.

\section{Results and discussion}

\subsection{Model evaluation}
\subsubsection{Extended winter means}
\begin{figure*}[h!]
	\centering
	\includegraphics[width=1\textwidth]{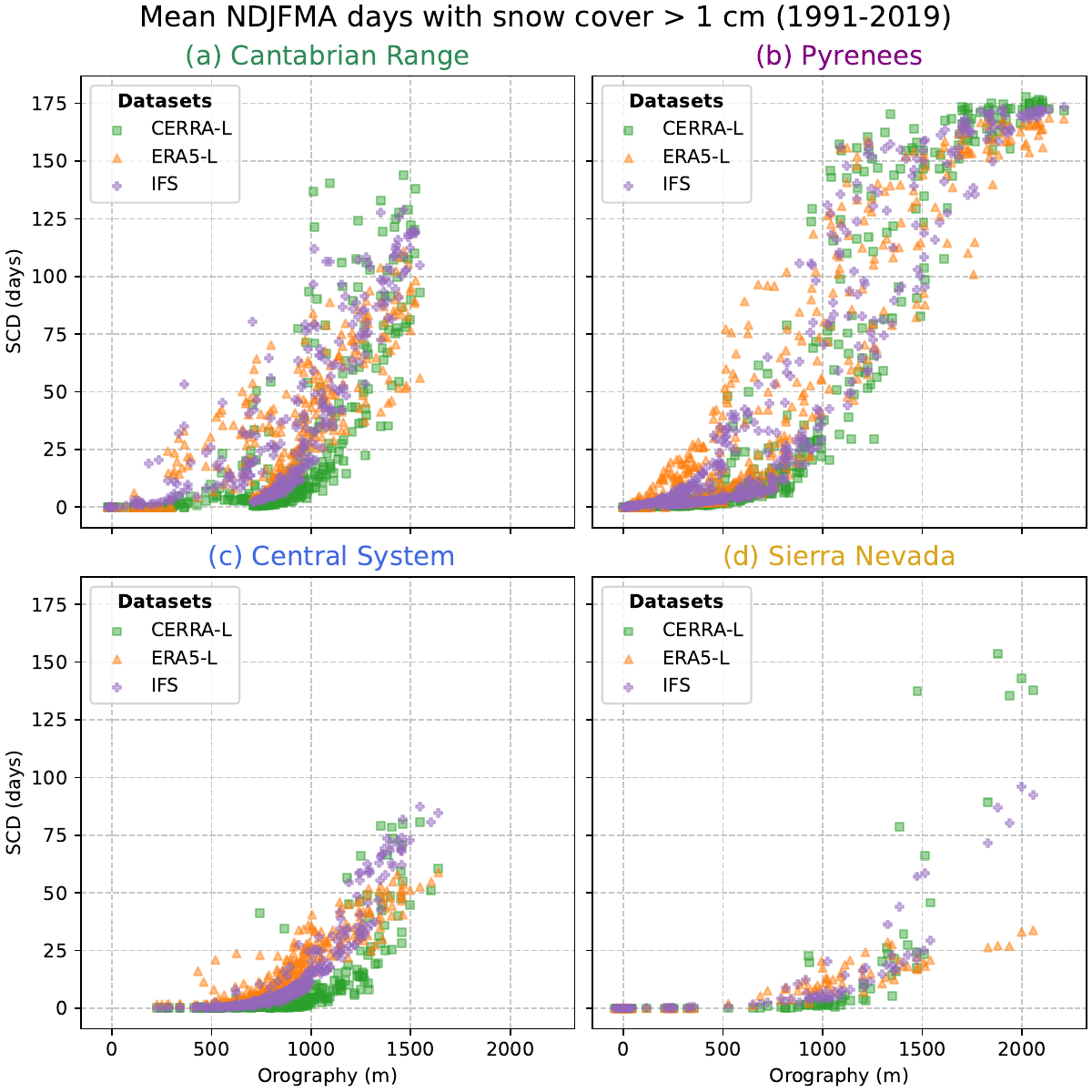}
	\caption{Mean number of days with a snow cover over 1 cm during NDJFMA versus grid point altitude. Each dot corresponds to a different grid point of the common 0.125º lat-lon grid. Each subplot corresponds to the grid points enclosed in the rectangular areas around each of the main mountain ranges, as shown in Figure \ref{fig: orog_ifs_zones}.}
	\label{fig: scd_by_altitude}
\end{figure*}

The length of the snow-covered season in the Iberian mountains is highly dependent on elevation and mountain range \citep{satelite_alonso-gonzalez_2020, gonzalez2022characterising, satelite_hidalgo_2024,melon-nava_snow_2025}. The IFS historical simulation captures the elevation dependency of the mean historical (1991-2019) number of NDJFMA days with a snow cover of more than 1 cm, and is mostly seen to be within the range of agreement with the main land surface reanalysis considered (Figure \ref{fig: scd_by_altitude}). Note that for this comparison, only CERRA-Land and ERA5-Land were represented alongside the IFS historical run to ensure the readability of the representation. These two datasets were chosen as two distinct and widely used reanalyses of the land surface. The IFS simulation outputs lie between those from ERA5-Land and CERRA-Land for most elevations in all four mountain areas. ERA5-Land renders more snow cover days at low to mid elevations and seems to present a more linear increase of this variable with elevation. Conversely, both CERRA-Land and the IFS historical run present a more sigmoid-like dependency with elevation, similar to previous works that used satellite and regional model data \citep{satelite_alonso-gonzalez_2020, satelite_hidalgo_2024,melon-nava_snow_2025}. This different behavior in ERA5-Land might be related to how the reanalysis was constructed, with a lower resolution ERA5 atmospheric forcing that could provide a smoothing effect on regions with narrow mountain systems such as the Iberian Peninsula. This may also explain the large differences observed at the higher elevations of the Sierra Nevada, where ERA5-Land appears unable to accurately represent the region, resulting in too few snow cover days. Considering this, the IFS historical run appears to more accurately capture the altitude dependency of snow cover duration than ERA5-Land. However, it still presents differences with respect to CERRA-Land as it generally simulates a higher mean number of SCD, specially at low- to mid- elevations. This is especially noticeable at the Central System, where the IFS presents slightly higher means than CERRA-Land at almost all elevations. This positive bias is also suggested at low- to mid- elevation in the Cantabrian range, where the spread between similar altitude grid points is also high in the IFS run. This positive bias can also be identified in low- elevations of the Pyrenees, while at higher altitudes both IFS and CERRA are found to be fairly consistent.  Conversely, in the Sierra Nevada, the IFS is largely consistent with CERRA-Land or even shows a negative bias at the highest elevations. This difference may be due to the smaller scale of this mountain range, which likely requires an even higher resolution to be fully resolved. 

\begin{figure*}[h!]
	\centering
	\includegraphics[width=1\textwidth]{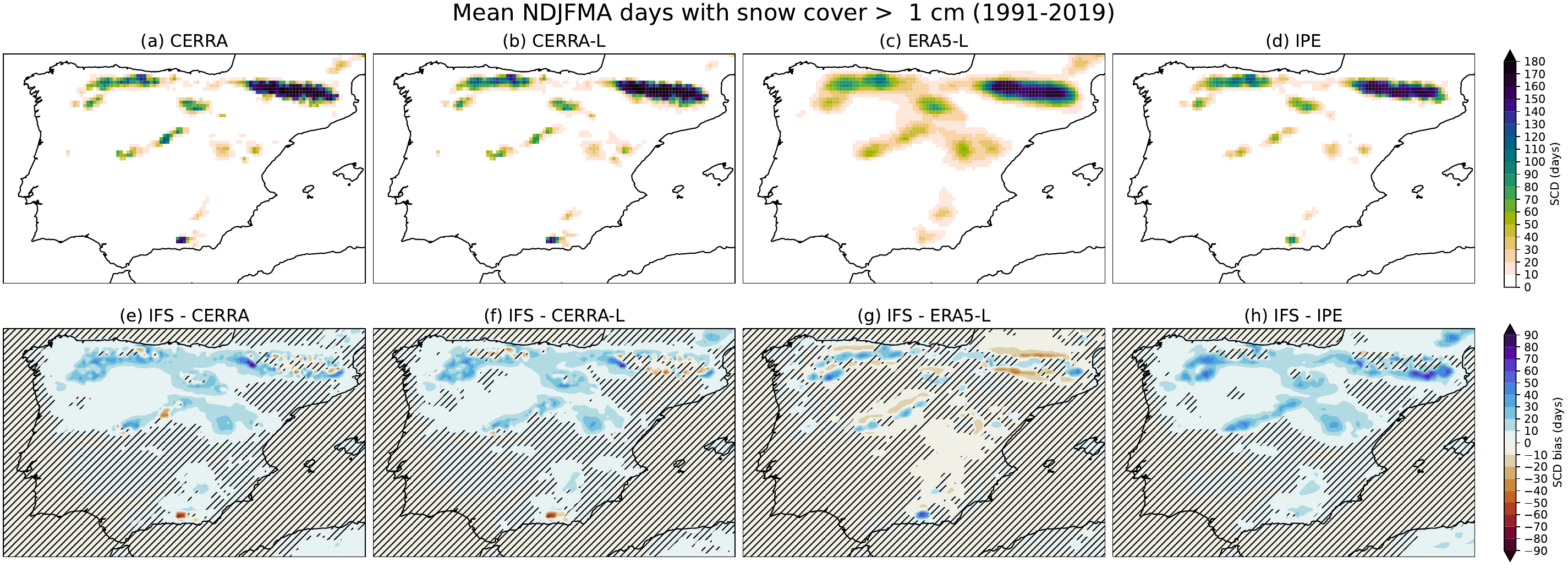}
	\caption{Maps of the mean number of days with a snow cover over 1 cm during NDJFMA for the reference datasets (a-d) and mean bias of the IFS historical simulation with respect to each reference dataset (e-h). The period considered is 1990 to 2019 (except for the IPE dataset which ends in 2014). Statistical insignificance ($\alpha_{FDR}=0.05$) of the bias is signaled by line hatching.}
	\label{fig: scd_datasets_bias_maps}
\end{figure*}

To better evaluate the potential positive bias in the mean snow cover duration, maps of both the climatology of SCD in each observation-derived dataset and the IFS model's bias relative to them are presented in Figure \ref{fig: scd_datasets_bias_maps}. The climatology for the IFS SCD for the same period can be found in Figure \ref{fig: intro_nextGEMS_vs_CMIP6}d. SCD mean climatologies are quite similar in CERRA and CERRA-Land, as expected given their shared atmospheric data. The IPE-CSIC dataset generally represents a slightly less extensive snow cover than the other datasets. In turn, ERA5-Land shows a smoother SCD climatology, displaying excessive snow cover days at low altitudes, rendering a too small and unrealistic altitude gradient, and an apparent larger region where snow cover is relevant. This differs from all other datasets and from the IFS historical run.

The overall sign of the SCD bias is positive (Figure \ref{fig: scd_datasets_bias_maps}, e-h). This indicates that the previously suggested positive bias in the length of the snow cover season is robust when compared with more datasets and is present at almost all zones. Despite this, spatial differences are identifiable in the bias maps against CERRA, CERRA-Land, and IPE-CSIC. Across nearly all major mountain ranges, and particularly in the Pyrenees and the Cantabrian range, the positive bias is more pronounced at low to mid elevations, decreasing notably at the highest elevations, where the bias becomes statistically insignificant. IFS agrees or even shows a negative bias when compared to ERA5-Land in some low-elevation areas, where the positive bias identified in the IFS and the smoothing identified in ERA5-Land might introduce similar errors. Note that the IFS exhibits an improved representation of the Sierra Nevada compared to ERA5-Land, while also being the only mountain range for which a clear negative bias is identified relative to CERRA and CERRA-Land. Sierra Nevada, being the smallest and southernmost of the studied mountain ranges while also containing the highest elevations of the Iberian Peninsula, might render it a special case where even higher resolution might be needed, but also where the positive effects of having a 9 km global coupled climate simulation are more noticeable. 

\begin{figure*}[h!]
	\centering
	\includegraphics[width=1\textwidth]{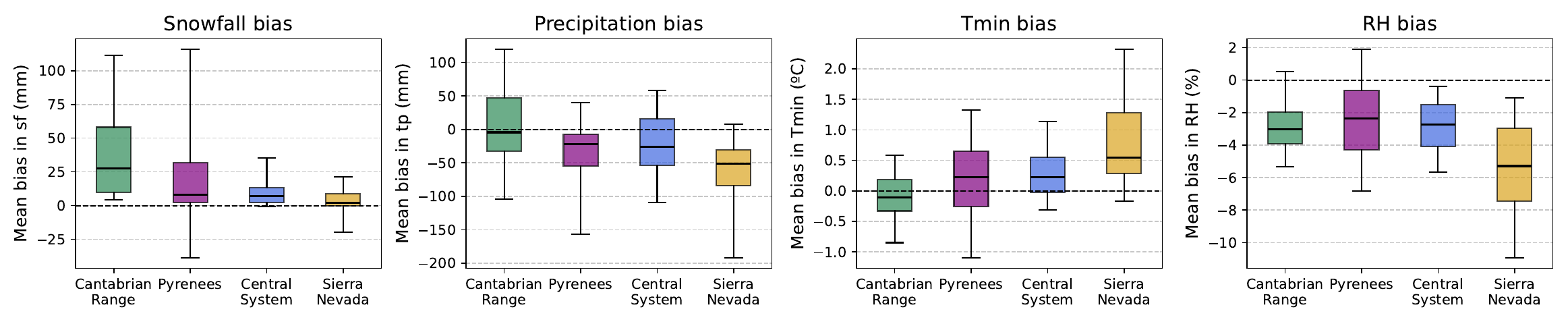}
	\caption{Spatial distribution of the mean NDJFMA bias of snowfall (sf), total precipitation (tp), minimum 2-meters temperature (Tmin), and 2-meters relative humidity (RH) for the main Iberian mountain ranges as defined in Figure \ref{fig: orog_ifs_zones}. The mean bias is calculated for each grid point with respect to the CERRA reanalysis and the results are shown as a boxplot. Only grid points over 700 m of altitude are considered.}
	\label{fig: bias_extra_variables_cerra}
\end{figure*} 

To identify possible sources of the positive bias in snow cover duration detected in the IFS, we compare IFS's total precipitation, total snowfall, minimum 2-meters temperature, and 2-meters relative humidity at the surface to CERRA reanalysis. In this case, CERRA was selected because it provides the most complete set of atmospheric variables among the reference datasets used in this work, while presenting a similar bias pattern for IFS in snow cover duration. The mean bias of IFS's atmospheric variables was calculated separately at each grid point above 700 m of altitude and represented as boxplots for the four mountain ranges to visualize its spatial variability (Figure \ref{fig: bias_extra_variables_cerra}). 
Figure \ref{fig: bias_extra_variables_cerra}a shows a positive bias in the mean total accumulated snowfall for all mountain ranges, with more deviation and spatial variability present in the Cantabrian range and the Pyrenees. This suggests that the previously identified positive bias in the duration of the snow-covered season is mostly due to excessive snowfall in the IFS compared to CERRA. The excess in snowfall could, in turn, be related to too much total rainfall and/or too cold surface temperatures. However, the spatial distribution of the mean bias in total precipitation shows no clear positive deviation in none of the studied mountain ranges (Figure \ref{fig: bias_extra_variables_cerra}b). Conversely, there is even a small negative bias in total precipitation in the Sierra Nevada and, to a lesser degree, in the Pyrenees. As for the minimum temperature bias (Figure \ref{fig: bias_extra_variables_cerra}c), there appears to be no clear sign for either the Cantabrian range or the Pyrenees, where the snowfall bias was larger. Moreover, in the Central System and especially in Sierra Nevada, there is a positive bias in the minimum temperature, which would decrease the likelihood of snowfall. We finally turn our attention to another variable that can influence the phase of precipitation: relative humidity. \cite{rh_nature} showed that snowfall can occur at higher temperatures when relative humidity is low, so a negative bias in relative humidity with respect to CERRA could contribute to the positive snowfall bias in the IFS. In fact, a relative humidity negative bias can be identified in all the studied mountain ranges (Figure \ref{fig: bias_extra_variables_cerra}d), with values systematically between 2 to 4 percentage points lower in the IFS than in CERRA. This dry bias could explain why more snowfall occurs in the IFS than in CERRA with similar amounts of total precipitation and similar ranges of surface temperature.

\subsubsection{Mean seasonal cycle}
\begin{figure*}[h!]
	\centering
	\includegraphics[width=1\textwidth]{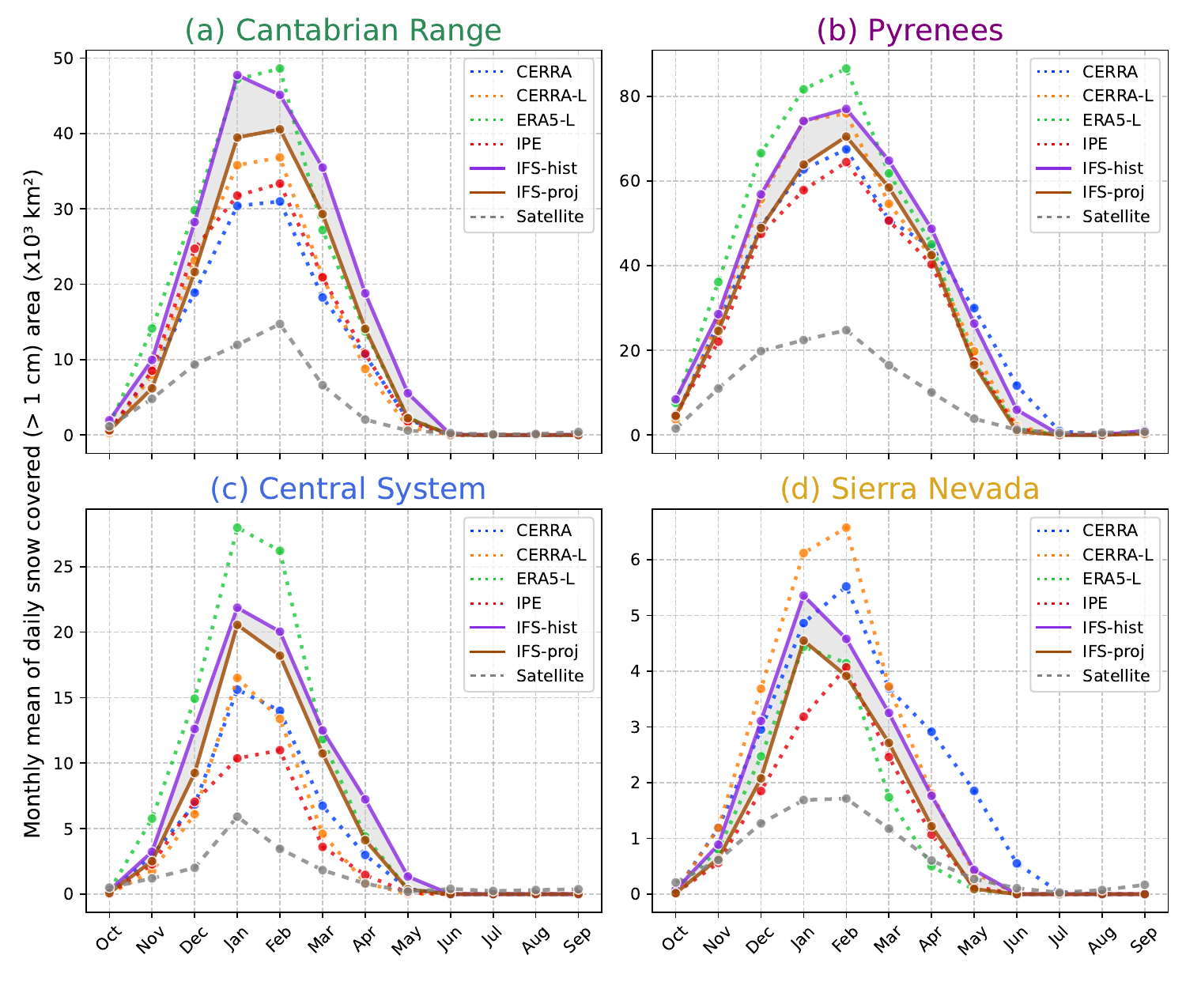}
	\caption{Seasonal cycle of monthly averaged snow covered area within each of the rectangular regions defined in Figure \ref{fig: orog_ifs_zones} for each mountain range of interest. The datasets and the IFS historical simulation consider the mean of each month for the period of 1990 to 2019 (except the IPE dataset which ends in 2014), while the IFS SSP3-7.0  simulation (2020-2049) is also displayed to allow the identification of climatic changes. MODIS satellite data are shown for the period 2003-2023.}
	\label{fig: sca_seasonal_cycle}
\end{figure*}

We now turn our attention to IFS's depiction of the snow season by computing the seasonal cycle using monthly  means of the snow-covered area (SCA) (Figure \ref{fig: sca_seasonal_cycle}). We perform this calculation for each dataset as geographical aggregate of each of the mountain ranges shown in Fig. \ref{fig: orog_ifs_zones}.

The snow season in the Iberian mountain ranges extends roughly from November to April, lasting longer in the Pyrenees and Sierra Nevada, where the highest altitudes of the peninsula are located (Fig. \ref{fig: orog_ifs_zones}). This corroborates the election of NDJFMA as the extended winter for this work. Regarding the different reference datasets, ERA5-Land usually behaves the most differently from the other three, showing higher values of snow-covered area, especially during the central months of winter, which is consistent with the previously discussed smoothing effect in this dataset. If the area that receives snow is smoothed, excessive snow will be present at lower altitudes, providing a larger mean snow-covered area. Regarding the IFS historical simulation, for most months and in all mountain ranges but Sierra Nevada, its values are above those from the reference datasets except ERA5-Land. This too large area is consistent with the previously identified positive bias in snow cover in low-to-mid altitudes. Despite differences in the total area, there is high agreement in the shape of the seasonal cycle among the datasets and the IFS runs. The peak months are similar in both the Pyrenees and the Central System, while a bit more uncertainty exists in the Cantabrian Range. There is more disagreement in Sierra Nevada, where ERA5-Land and the IFS highlight January as the peak month, while the other datasets indicate February. It is also interesting to note that the IFS seems to render a wider shape for the seasonal cycle, with larger areas during the spring months, especially at the Central System and the Cantabrian Range. 

The MODIS satellite observations are also included in Figure \ref{fig: sca_seasonal_cycle}. It should be noted that for the MODIS data, its maximum resolution (0.005 $^\circ$) was used to compute the SCA throughout all analyses in this study. As so, it is to be expected that the mean values of SCA obtained are considerably lower than those computed with either of the other datasets from the common 0.125 $^\circ$ resolution. The decision to use MODIS data at its maximum resolution was made in order to remark the limitations still present at kilometer-scale simulations and reanalysis data when compared to sub-kilometer observational data. Nevertheless, MODIS data can also be used here to verify the shape of the seasonal cycle with direct observations. For the Pyrenees and the Central System, the satellite is in line with the IFS and most of the datasets regarding the peak month and overall shape. In Sierra Nevada, satellite data indicates a peak between January and February, in line with the uncertainty presented by the datasets between those two months. For the Cantabrian range, MODIS shows a delayed peak in February in agreement with just some of the datasets and shifted with respect to the IFS historical simulation.

\subsubsection{Distribution and interannual variability }
\begin{figure*}[h!]
	\centering
	\includegraphics[width=1\textwidth]{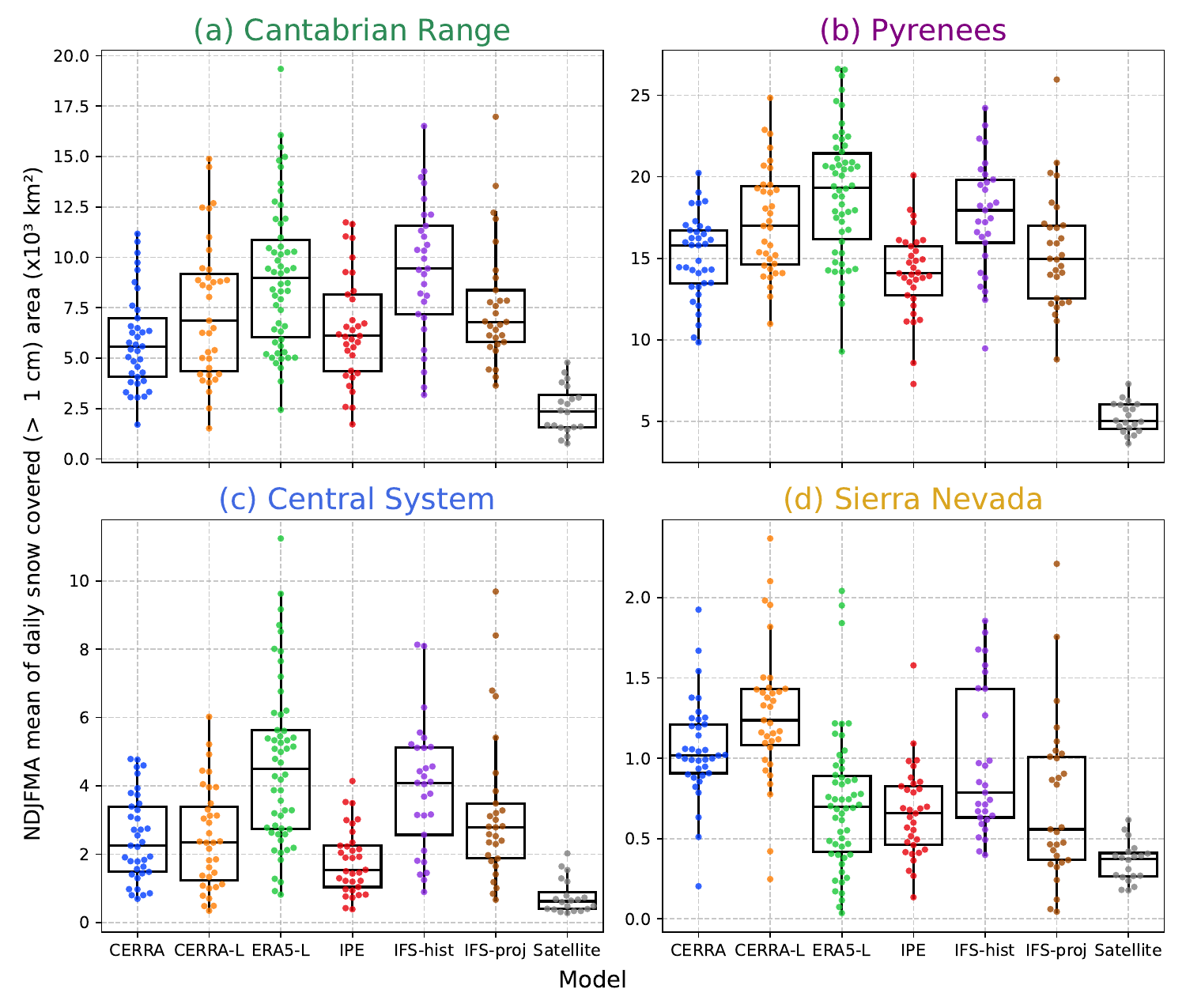}
	\caption{Boxplots of snow-covered area averaged during NDJFMA within each of the rectangular regions defined in Figure \ref{fig: orog_ifs_zones} for each mountain range of interest. Each colored dot represents the mean value of one single extended winter within the studied period. The datasets and IFS-historical consider the period of 1990 to 2019 (except for the IPE dataset, which ends in 2014). The IFS SSP3-7.0 simulation (2020-2049) is also displayed to allow the identification of climatic changes. MODIS satellite data are shown for the period 2003-2023.}
	\label{fig: sca_boxes}
\end{figure*}

We now focus on the interannual distribution of the snow-covered area averaged for entire NDJFMA extended winters. Each winter is represented for each dataset and mountain range, and the main statistical features of the distribution are highlighted via boxes and whiskers (Figure \ref{fig: sca_boxes}). The comparison suggests that for all mountain ranges except Sierra Nevada, the IFS historical climatology is in better agreement with ERA5-Land than with the other reference datasets. This was already highlighted when analysing the seasonal cycle (Figure \ref{fig: sca_seasonal_cycle}) and argued to be caused by the smoothing bias present in ERA5-Land and the excessive snow in IFS at low altitudes, both increasing the values of the snow-covered area. In terms of the variability, CERRA-Land, ERA5-Land, and the IFS historical run agree on the interquartile range, while CERRA and the IPE-CSIC dataset often present a narrower distribution. For the case of Sierra Nevada, the IFS presents a stretched distribution that may hint at a bimodality. This bimodality is not clearly evident in the reference datasets, aside from hints of a few extreme values in CERRA-Land and ERA5-Land. Furthermore, the MODIS satellite data does not have any bimodality nor extreme values, suggesting an overestimation of these in the models and some reanalysis. Extreme values similar to those that stretch the IFS distribution are present in all other datasets, just not enough to be considered more than outliers. 

\begin{figure*}[h!]
	\centering
	\includegraphics[width=0.97\textwidth]{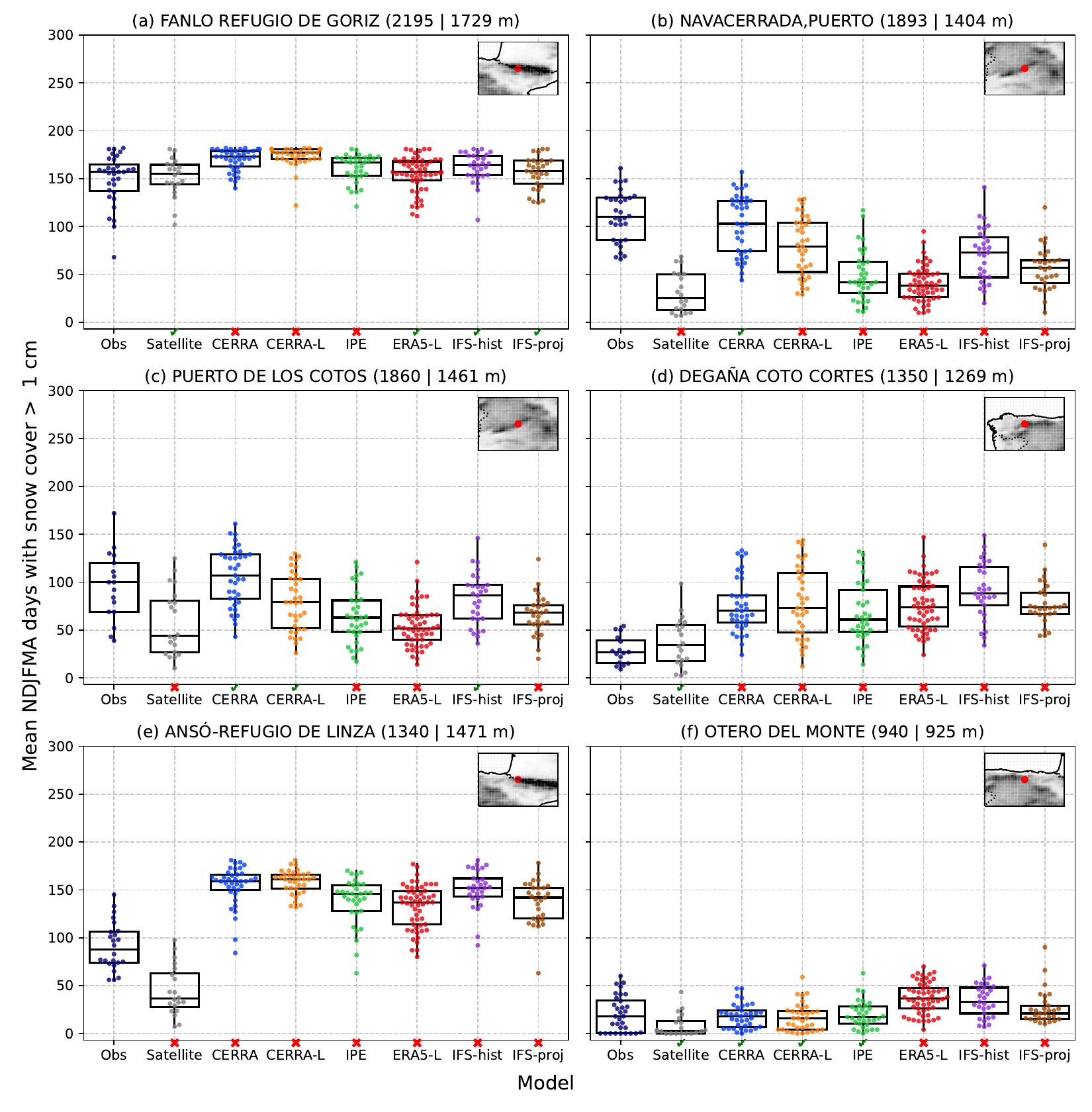}
	\caption{Comparison of model simulations, gridded datasets and satellite data with in situ observations. The nearest grid point to the observation sites (located at the red dots in the inset maps) has been considered in each case. Boxplots display the snow-covered days (SCD) during NDJFMA. Each colored dot represents the mean for one single extended winter within the studied period. The datasets and IFS-historical consider the period of 1991 to 2019 (2014 for IPE). MODIS satellite data are shown for the period 2003-2023. The observation sites measured period varies from one another, but each site has at least 15 observed winters within 1990-2019. The IFS SSP3-7.0  simulation (2020-2049) is also displayed. In each subplot, above the datasets names, a cross or a tick indicates whether the product's distribution is compatible or not ($\alpha=0.05$) with observations following a two-sample Kolmogorov-Smirnov test \citep{hodges_kolmogorob_1958}. Next to the name of each location, the altitude of the site is displayed first, followed by the altitude of the model nearest grid point.}
	\label{fig: scd_observations_boxes}
\end{figure*}

To examine the distributions and interannual variability of NDJFMA snow-covered days, we now focus on specific locations where manual snow depth observations exist, despite the overall scarcity of such measurements in the Iberian Peninsula. This approach also serves to evaluate the maximum capacity of kilometer-scale climate models and regional datasets at the local scale. It must be kept in mind that a model's grid point is meant to represent the mean of an extended area, so it will not necessarily agree with observations that sample a single discrete spatial point in the real world affected by many local factors not considered in the models. This fact is especially true for the case of snow depth, given its high spatial variability driven by highly complex processes such as orographic precipitation, preferential deposition, wind and gravitational redistribution or shading and sheltering \citep{snow_distrib_heter_1991,snow_distrib_heter_1992,snow_distrib_het_2011,snow_distrib_het_review_2011,HICARsnow}. 

The number of NDJFMA days with a snow cover of more than 1 cm is calculated for each winter, site, and dataset and displayed as dots in Fig. \ref{fig: scd_observations_boxes}. The distribution is shown with boxplots, showcasing the main statistical characteristics. The MODIS data was calculated using its maximum resolution of 0.005$^\circ$ in order to be as close to the point observations as possible. In general, there is consistency among the distributions of the reference gridded datasets and the IFS historical run. Moreover, for Fanlo Refugio de Goriz, there is great agreement between the local observation, satellite, reanalysis-based products and the IFS historical simulation, despite the inherent difficulty of comparing extensive grid point representation to point observations. It is relevant to note that the Fanlo site is in an exposed north-facing non-forested area, ideal for satellite measurements. There is also good agreement for the Puerto de los Cotos site between observations, CERRA and the IFS. In this case, the satellite might be slightly underestimating when compared to point observations due to the presence of forested and heavily human-influenced terrain around the site's location.    
However, at other sites such as Degaña Coto Cortes or Ansó-Refugio de Linza, despite the consistency shown by the reference datasets and the IFS model, all presenting similar means, the point observations fall widely behind, maybe indicating that the measurement site lies in a location not representative enough of the extended zone averaged by the 0.125$^\circ$ grid point in the reanalysis-based products. In the case of Navacerrada, there is a strong dispersion among the different datasets, with CERRA, CERRA-Land, and the IFS historical being the closest to the observations. There, the satellite significantly underestimates the duration of the snow season due to the site being located inside a heavily forested south-facing area, prone to shadows difficulting satellital snow detection. The case of Otero del Monte might hint at the smoothing bias and the positive bias in low to mid elevations already discussed for ERA5-Land and the IFS, respectively, as they show larger values than expected for a low-altitude zone, while all other datasets are in line with the observations. In this case the satellite is mostly in line with observations while still underestimating slightly. 

Overall, differences exist between the reference datasets, the model simulation, and the observations. Our results suggest that the IFS presents a positive bias in snow cover days at low- to mid- elevations, which can be traced to excessive snowfall, probably linked to a negative bias in relative humidity, which leads to an overestimation of precipitation falling as snow. This positive bias affects the estimates of total snow-cover area (Figures \ref{fig: sca_seasonal_cycle} and \ref{fig: sca_boxes}), for which the IFS historical simulation was found to exhibit a positive bias both in the mean NDJFMA climatology and in each month of the annual cycle. However, we note that these climatologies were still within the range of some of the reference datasets, and even compatible with point observations at some sites (Figure \ref{fig: scd_observations_boxes}). These results are encouraging, especially given the challenges associated with comparing gridded data to local measurements. These findings suggest that global km-scale climate model runs, such as the nextGEMS simulations here analysed and future initiatives like Destination Earth \citep{destinationearth2021,doblas-reyes_destination_2025}, could be used to assess the future evolution of local phenomena in a changing climate, enabling more targeted studies for adaptation purposes. 

\subsection{Future changes}
\subsubsection{Changes in the snow cover duration}
Once the capabilities and biases of the IFS model in representing various aspects of the climatology and variability of seasonal snow cover in the main Iberian mountain regions have been established, the SSP3-7.0 projection can be compared to the historical run to assess the projected impacts of climate change in these sensitive areas. 

For the SSP3-7.0 projection, the mean of the daily snow-covered area decreases both in its whole extended winter climatological distribution (Figure \ref{fig: sca_boxes}) and for each month (Figure \ref{fig: sca_seasonal_cycle}). At the monthly timescale, the reduction in snow-cover area is mostly uniform across the seasonal cycle. The only exception is the Cantabrian range, where the peak month shifts from January in the historical simulation to February in the projection. We note, nevertheless, that there was no clear agreement between the reference datasets regarding the month of occurrence for the maximum. The extended winter climatological mean decreases in all mountain ranges, and the interannual variability is also somewhat reduced in all regions but the Pyrenees (Figure \ref{fig: sca_boxes}). Nonetheless, even with said reduced mean and variance, there are still outliers with exceptionally large mean snow-covered areas for all mountain ranges in the projection. Some of these outliers are even the highest values of this variable when taking into account both the historical and projection runs, and in the Sierra Nevada and the Pyrenees, the outliers present the highest values across all the considered datasets. Other features of the historical simulation, such as the apparently bimodal distribution of SCA in Sierra Nevada, are still present in the projected climatology. It is also worth noting that the projected climatologies sometimes show closer agreement with certain reference datasets than the historical simulation. However, this may result from a compensation of differences, where biases in either the historical simulation or the reference datasets are partially offset by the climate change–driven decrease in snow. 

Consistent with the decrease in the snow-covered area, the number of days with a snow cover greater than 1 cm is also projected to decrease, even at the local scale of the grid point closest to the observation sites (Figure \ref{fig: scd_observations_boxes}). In all six studied sites, the distribution of snow-covered days during NDJFMA is shown to shift to lower values, and in most cases, it is also shown to narrow, indicating a reduction of variance. Again, compensating errors might render cases where the projected distribution is more in line with the observations than the historical run, but these do not appear to happen systematically.

\begin{figure*}[h!]
	\centering
	\includegraphics[width=1\textwidth]{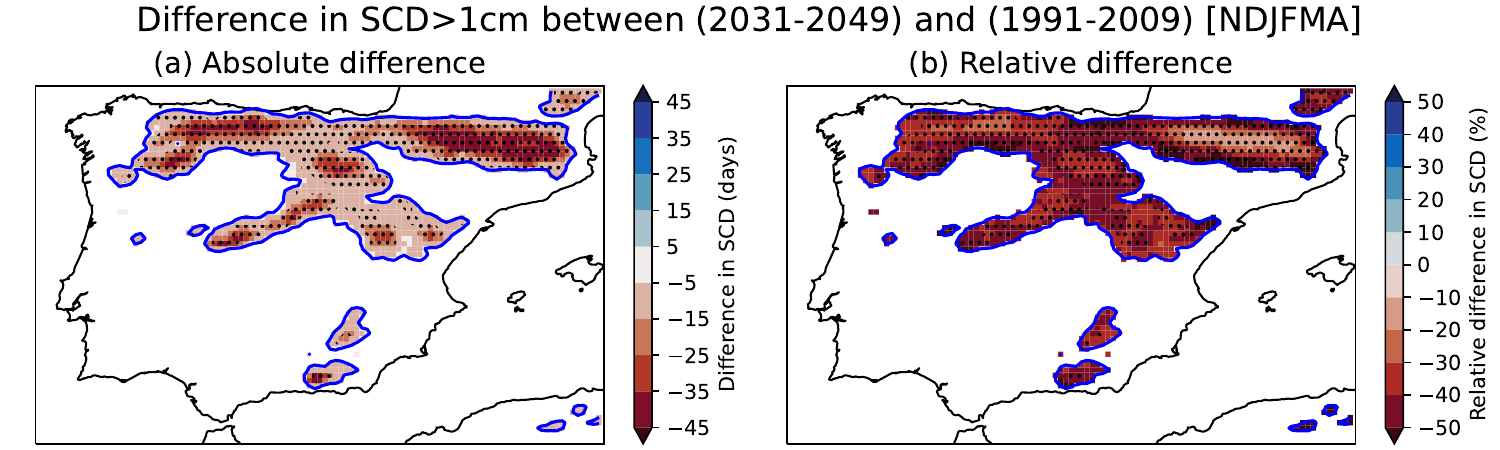}
	\caption{Maps of the absolute (a) and relative (b) change in Snow Cover Days between the last 19 extended winters (2031-2049) of the SSP3-7.0 projection and the first 19 extended winters (1991-2009) of the historical run. Relative values are computed with respect to the first 19 extended winter means. Only grid points with at least 10 SCD in the 1991-2009 mean are considered and they are also highlighted by a blue contour line. Stippling signals statistical significance ($\alpha_{FDR}=0.05$).}
	\label{fig: change_scd_maps}
\end{figure*}

To evaluate whether the projected snow-cover reduction shown at the selected sites is part of a more general feature, we now focus on the changes across all mountain grid points. We choose to compare two parts of the historical and SSP3-7.0 simulations, long enough to provide meaningful statistics but also distant enough so that the external forcing is different. Those periods are the first 19 extended winters of the historical run (1991-2009) and the last 19 ones of the projection (2031-2049). The difference of the mean number of NDJFMA days with snow cover over 1 cm is calculated between both periods, showing a decrease at all mountain grid points (Figure \ref{fig: change_scd_maps}). This decrease is also statistically significant at almost all mountain grid points, the only exception being the lowest elevations where the standard deviation is high (not shown) compared to a low mean number of SCD (see Figure \ref{fig: intro_nextGEMS_vs_CMIP6}d). The reduction of snow cover days is stronger in absolute terms in those regions where the climatological number of SCD is higher (as was shown in Figure \ref{fig: intro_nextGEMS_vs_CMIP6}d and the upper raw of Figure \ref{fig: scd_datasets_bias_maps}), those being mostly the highest areas of each mountain range. However, when relative differences are considered, the changes are more homogeneous while still exhibiting some spatial variability. On the one hand, there is an elevation gradient with lower altitude zones losing up to more than 50 percent of their snow cover days. On the other hand, there is a north-to-south gradient, which can be seen especially over Sierra Nevada, where its highest elevations lose more than 40 percent of their SCD, versus the Pyrenees, where its highest points lose less than 20 percent. 

\subsubsection{Drivers of the reduction in snow cover duration}
\begin{figure*}[h!]
	\centering
	\includegraphics[width=1\textwidth]{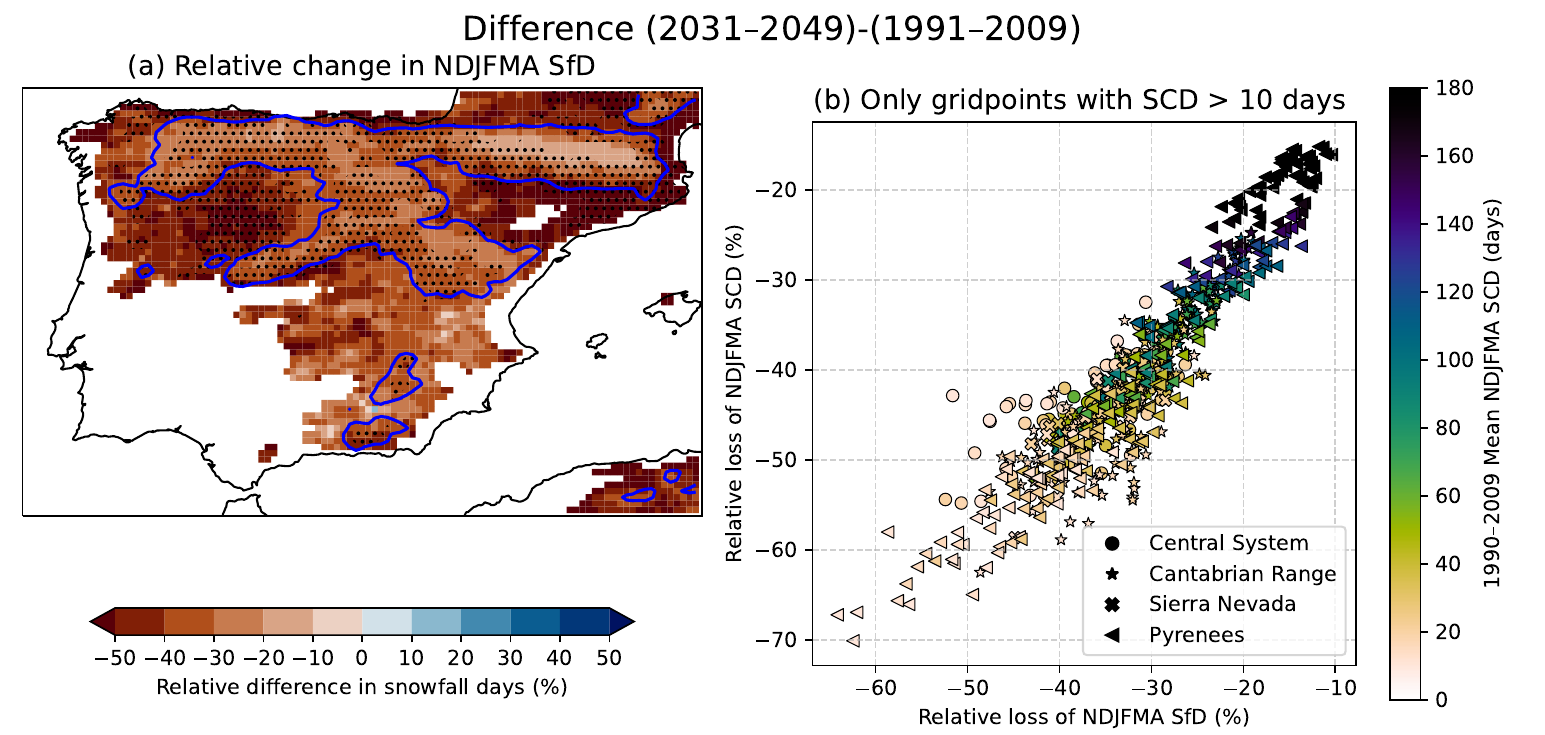}
	\caption{Relationship between changes in snow cover days and changes in snowfall days. (a): map of the relative change in snowfall days observed between 2031-2049 and 1991-2009. Stippling signals statistical significance ($\alpha_{FDR}=0.05$). (b): scatter-plot of the relative change of SCD versus the relative change of SfD for each gridpoint of the four mountain regions of interest. Grid points with at least 10 SCD in the 1991-2009 mean are highlighted by the blue contours in the map and are the only grid points considered for the scatter.}
	\label{fig: sfd_scd_scatter}
\end{figure*}

The loss of snow cover days could be associated with many different drivers \citep{snow_drivers_2009,snow_drivers_2015,alonso-gonzalez_snowpack_sensitivity_2020}; however, the most immediate one would be a reduction in snowfall days caused by either a rise in temperature or a decrease in precipitation. Another process that affects the duration of the snowpack season and may be influenced by changes in temperature and precipitation patterns is snow-melt. Temperature rise may increase melting rates and increase the frequency of rain-on-snow events that can cause rapid melting extremes \citep{snow-in-cc_2005,rain-on-snow_2016,rain-on-snow_2018,bonsoms_rain--snow_2024}. To study some of these possible drivers, the relative changes in the number of snowfall days have been computed. Figure \ref{fig: sfd_scd_scatter}a shows that snowfall days also decrease across all grid points, with the reduction being statistically significant in almost all mountain regions. Furthermore, there is a clear spatial correlation between the relative loss of snow cover days and that of snowfall days (Figure \ref{fig: sfd_scd_scatter}b). Spatially, grid points that experienced a greater relative reduction in snow cover days are also those most affected by the relative decrease in snowfall days. The temporal correlation of these two variables was also computed considering for each gridpoint both historical and projection runs and found to have a median Pearson correlation coefficient of 0.81 with an inter-quartile range of 0.10, indicating that the changes in snowfall days may be the most influential factor to the changes in the duration of the snow cover season. Following this, further analysis will focus on finding whether this reduction in snowfall days might be more related to precipitation or to temperature changes in the different regions. The possible changes in melt rates will not be explored in this work.

\begin{figure*}[h!]
	\centering
	\includegraphics[width=1\textwidth]{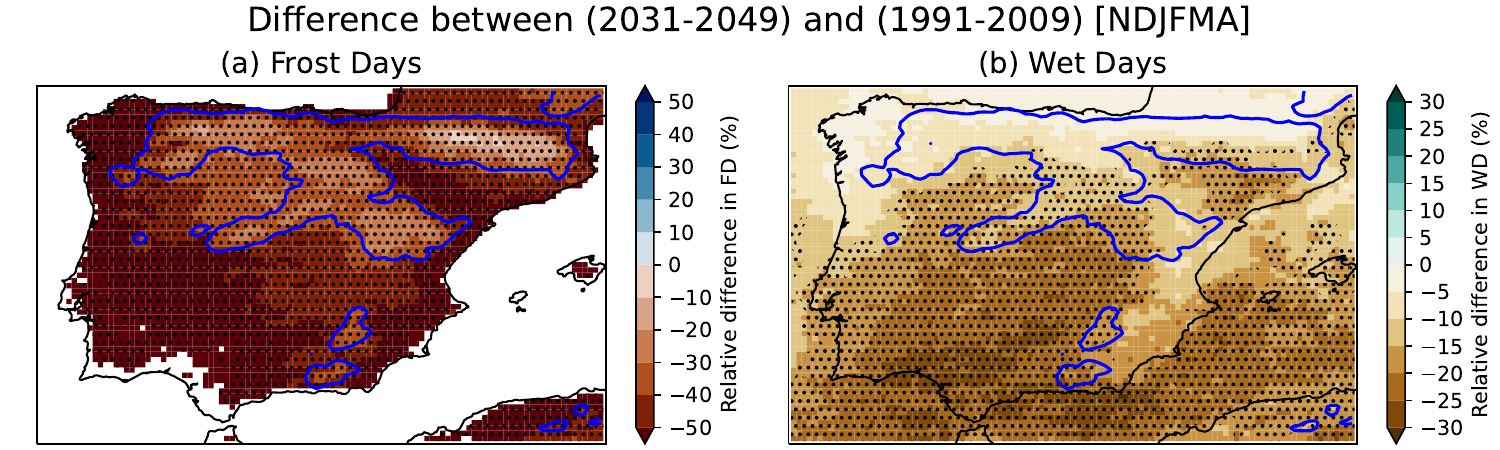}
	\caption{Same as Figure \ref{fig: change_scd_maps} but for the relative changes in Frost Days (a) Wet Days (b).}
	\label{fig: change_wd_fd_maps}
\end{figure*}

To determine the influence of temperature and precipitation changes on snowfall days, new indices are computed. The number of snowfall days can be highly related to the number of wet days and the number of days where freezing temperatures are reached, also known as frost days. These two indices have also been successfully used to infer the number of potential snowfall days when this data was not available and found to be highly accurate when compared to the true snowfall days \citep{duran_analysis_2024,duran_multidecadal_2024}. This compound index of potential snowfall days has also been compared to a true snowfall days index within the IFS model historical simulation and found to be in good agreement with per gridpoint Pearson correlation coefficient between both indexes aggregated for full NDJFMA winters having a median of 0.89 an an inter-quartile range of 0.08. As so, changes in either of these indices should be closely related to changes in snowfall days, while also offering the possibility of uncoupling the different influences of precipitation and temperature changes in the simulations. These changes have been computed (Figure \ref{fig: change_wd_fd_maps}) and found to be statistically significant in extended parts of the studied mountain regions. For the case of frost days, a relative reduction is found in all of the Iberian Peninsula, though this relative reduction seems to have an altitude gradient, where it is less intense at higher elevations. In the case of wet days, the sign of the changes is also negative for all of the Iberian Peninsula. However, changes are not statistically significant at all grid points, and a clear north-south gradient can be seen, where the number of precipitation days is reduced more intensely at lower latitudes and on the southern slopes of the main mountain ranges. 

\begin{figure*}[h!]
	\centering
	\includegraphics[width=1\textwidth]{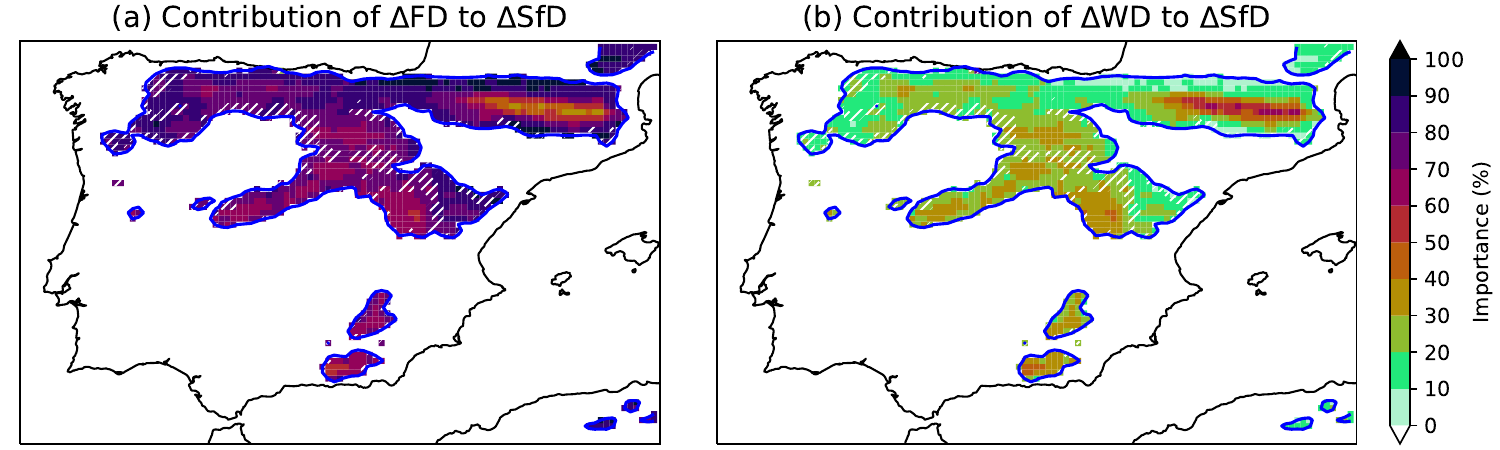}
	\caption{Relative contribution of the change in Frost Days (a) and the change in Wet Days (b) to the change in Snowfall Days between the last 19 extended winters (2031-2049) of the SSP3-7.0 projection and the first 19 extended winters (1991-2009) of the historical run. The importance of each term is computed considering a multiple linear regression model. The percentage displayed in each sub-figure corresponds to the regression coefficient multiplied by the observed change in each variable and divided by the approximate change in Snowfall Days obtained with the multiple linear regression, thus both maps are complementary. Regions where the variance of the residual is greater than 0.5 of the total variance of SfD are marked with lines and are not considered useful for analysis.}
	\label{fig: regress_contributions}
\end{figure*}

Both the reduction in frost days and that in wet days can be driving the projected decrease in snowfall days. However, the contribution of each term might depend on the mountain range and local environment. To quantify this contribution, a multiple linear regression analysis was performed (see section \ref{sec: Methods_reg} and \cite{libro_estadistica_Storch}). The objective was to generate a simple model that could approximate the changes in snowfall days by a linear combination of the changes in wet days and frost days. The linear regression was trained at each grid point ($k$) with the anomalies at interannual timescales, providing the coefficients ($a$ and $b$) to the following expression: $ \mathrm{\Delta SfD}_k = a \mathrm{\Delta FD}_k + b \mathrm{\Delta WD}_k +\epsilon_k$, where $\epsilon_k$ represents the residual of the fitting. To increase the sample size and account for behavior changes due to climate change, the annual SfD, WD, and FD anomalies of both the historical and the projection runs where considered.  A series of statisticals where computed in order to asses the performance of the model. The coefficient of determination ($R^2$) was calculated for each gridpoint and found to have a median value of 0.60 and an inter-quartile range of 0.21. The covariance term was also computed for each gridpoint and found to account for less than 20\% of the total variance in 91\% of the gridpoint. 

Once the linear regression limitations have been assessed, it can be concluded that it serves as a first approach to disentangle the effects of precipitation and temperature changes on snowfall. As so, using this simple per-gridpoint model, the change in snowfall days ($a\Delta \mathrm{SfD}$) between the 1991-2009 and the 2031-2049 climatologies can be approximated as a sum of the coefficient multiplied by the change in frost days ($\Delta \mathrm{FD}$) and wet days ($b\Delta \mathrm{WD}$). To assess the validity of this approximation, the estimated values were compared with the changes in SfD (as computed in Figure \ref{fig: sfd_scd_scatter}a). The results (not shown) indicate that the approximation closely reproduces the changes, although it systematically overestimates by approximately one to two days in most cases.

Following this approach, the contribution of the change in frost days can be approximated as: $\frac{a\Delta \mathrm{FD}}{\Delta \mathrm{SfD}}\times100$. This quantity, as well as the corresponding one for wet days, can be used to provide a simple spatial representation of the contributions of each variable change to snow (Figure \ref{fig: regress_contributions}). The analysis suggests that the main driver of the decrease in snowfall days is the temperature rise. There is, however, some spatial variability present between and within the main mountain ranges. The reduction of frost days appears to be more important at low to mid elevations, where a temperature rise might vastly reduce the chance of snowfall. At the highest elevations, where there are more days when snowfall is possible, the reduction of precipitation also starts to play a role. This role becomes more important at the southern mountain ranges, in the Central System, and more clearly Sierra Nevada, where a significant decrease in precipitation days was present in the IFS simulations (Figure \ref{fig: change_wd_fd_maps}). 

Overall, the length of the snow-covered season has been found to decrease significantly in all mountain ranges between the historical 1991-2009 period and the projected 2031-2049 period. However, this reduction is not homogeneous and instead presents more relative impact in low- and mid-elevations as well as in the southernmost areas (Figure \ref{fig: change_scd_maps}). The main immediate cause of this reduction was found to be a decrease in snowfall days (Figure \ref{fig: sfd_scd_scatter}). This is driven both by a temperature rise and a reduction of wet days, though the contribution of these factors depends on altitude and geographical location (Figure \ref{fig: regress_contributions}). For the Cantabrian Range, the main driver is  temperature rise as wet days are not found to change significantly (Figure \ref{fig: change_wd_fd_maps}). In the case of the Sierra Nevada and the Central System, the temperature rise is also dominant but a significant reduction in precipitation also contributes. Finally, in the Pyrenees the reduction in wet days is found to be the main driver in high-elevations and in mid-elevations of its southern side (Figure \ref{fig: change_wd_fd_maps}) while the reduction of frost days is key in its northern side and low-elevations.

\section{Conclusions}
Mountain seasonal snow cover is a vital hydrological resource for Mediterranean regions such as the Iberian Peninsula. As climate change is expected to affect mountain snow cover, climate projections that can provide quality information at regional to local scales became necessary for informed adaptation efforts. Previously, the coarse resolution of global climate models has limited and delayed the availability of such information, specially regarding variables heavily influenced by orographic processes such as mountain snow depth. In this regard, recent modeling efforts in global kilometer-scale simulations such as the nextGEMS-EU project open up new possibilities. 

In this study we validate the nextGEMS IFS-FESOM model in terms of its representation of the seasonal snow cover in four main Iberian mountain ranges. For this purpose a 1990 to 2019 historical simulation is used. The validation was performed with respect to four different reanalysis-based datasets: CERRA, CERRA-Land, ERA5-Land and the IPE-CSIC gridded dataset. Manual snow depth observations and satellite data were also used. Our results suggest that IFS presents a positive bias in the number of days with a minimum snow cover depth of 1 cm at low- to mid- elevations in almost all mountain ranges (see Figure \ref{fig: scd_datasets_bias_maps} e-h). This positive bias is also present when analyzing the mean snow-covered area (Figures \ref{fig: sca_seasonal_cycle} and \ref{fig: sca_boxes}). It can be traced to an excessive proportion of the total precipitation falling as snow which, in turn, may be related to a negative bias in relative humidity (Figure \ref{fig: bias_extra_variables_cerra}). Despite this positive bias, the IFS is shown to be within the range of uncertainty provided by the different reference datasets in most zones. Even some local comparisons with in situ observations were shown to be compatible with the IFS (Figure \ref{fig: scd_observations_boxes}), despite the challenges associated with comparing extensive grid-point data to point observations. These findings suggest that global km-scale climate model runs could be used to assess the future evolution of local phenomena in a changing climate, enabling more targeted studies for adaptation purposes. 

Once the model is deemed capable of simulating the main historical climatology of the seasonal snow cover in the Iberian mountains, differences between the historical simulation and a 2020 to 2049 SSP3-7.0 projection are assessed. The length of the snow-covered season was found to decrease significantly in all mountain ranges between the periods 1991-2009 and 2031 to 2049. This reduction was found to be more pronounced in low- to mid- elevations and in the southernmost mountain areas (Figure \ref{fig: change_scd_maps}). For example, in terms of relative loss of snow covered days, most low elevations present reductions of more than 50\% of their initial climatologies while the highest points in the Pyrenees report the lesser changes at about 20\% reduction from their initial values. This reduction of the snow covered season was found to be closely related to a decrease in snowfall days (Figure \ref{fig: sfd_scd_scatter}) which, in turn, is caused by both an increase in temperature and a decrease in wet days (Figure \ref{fig: change_wd_fd_maps}). The contribution of this to drivers was found to be dependent on both altitude and latitude with the reduction in total precipitation days being more important at high elevations were the temperature is low throughout the winter and at the southern zones of the peninsula (Figure \ref{fig: regress_contributions}). 

Overall, despite the presence of model biases, nextGEMS IFS-FESOM was found to be able to represent the main climatological features of the seasonal snowpack in Iberian mountains. This has allowed to regionally asses near future projected changes of climatologies, showing a marked decline in snow cover duration especially affecting low- to mid- elevations and southern mountain ranges. We conclude that kilometer-scale global climate simulations can be used to fairly assess climate change impacts in specific regional phenomena such as seasonal snowpack in medium-sized Mediterranean mountains such as the ones in Iberia without the time delays, resources and potential inconsistencies associated to downscaling efforts, although such efforts will remain useful when dealing with local snow phenomena at sub-kilometric scales. 

\section{CRediT authorship contribution statement}
\textbf{Diego García-Maroto:} Writing – review and editing, Writing – original draft, Visualization, Software, Methodology, Investigation, Formal analysis, Data curation, Conceptualization. \textbf{Elsa Mohino:} Writing – review and editing, Supervision, Resources, Project administration, Methodology, Funding acquisition, Conceptualization. \textbf{Luis Durán:} Writing – review and editing, Supervision, Resources, Project administration, Methodology, Funding acquisition, Conceptualization. \textbf{Álvaro Gonzaléz-Cervera:} Writing – review and editing, Software, Methodology, Data curation. \textbf{Xabier Pedruzo-Bagazgoitia:} Writing – review and editing, Software, Data curation.

\section{Declaration of competing interest}
The authors declare that they have no known competing financial interests or personal relationships that could have appeared to influence the work reported in this paper.

\section{Acknowledgements}
We would like to thank the Agencia Estatal de Meteorología (AEMET) and specially Esteban Rodriguéz-Guisado and Juan Carlos Sanchez Perrino for providing observational data. 
\textbf{Funding:} This research was supported by the EU Horizon 2020 project nextGEMS under grant agreement no. 101003470. This research was also supported by the Spanish Ministry for Science and Innovation thorugh project FIRN (PID2022-140690OA-I00, Proyectos de Generación de Conocimiento 2022) and project DISTROPIA (PID2021-125806NB-I00,  Proyectos de Generación de Conocimiento 2021). DGM acknowledges financial support from the Spanish Ministry of Science, Innovation and Universities through the FPU23 fellowship (AP-2023-02964).

\section{Data availability}
The simulation data are openly accessible and archived in the World Data Center for Climate at \url{https://doi.org/10.35095/WDCC/next GEMS_prod_addinfov1} \citep{zenodo_nextgems}. Namelist files and settings for the 30-year Cycle 4 production simulation with IFS-FESOM are archived on Zenodo at \url{https://doi.org/10.5281/zenodo.14725225} \citep{ifs-nextgems}. ERA5-Land, CERRA, and CERRA-Land datasets are available from the Copernicus Climate Data Store (\url{https://cds.climate.copernicus.eu}). \cite{ipe} snow depth dataset is available on Zenodo at \url{https://doi.org/10.5281/zenodo.854619}. The MODIS Normalized Difference Snow Index (NDSI) MYD10A1 (Aqua) product is available from NASA’s National Snow and Ice Data Center (NSIDC) at \url{https://doi.org/10.5067/MODIS/MYD10A1.061}. AEMET snow depth observations were provided upon request and are not publicly available.

\bibliographystyle{elsarticle-harv} 
\bibliography{biblio}






\end{document}